\documentclass[preprint]{aastex}

\shorttitle{Temperature \& Helium Determinations}
\shortauthors{Peimbert, Peimbert, & Luridiana}

\begin{document}

\title{Temperature Bias and the Primordial Helium Abundance Determination}

\author{Antonio Peimbert, Manuel Peimbert, \& Valentina Luridiana}
\affil{Instituto de Astronom\'{\i}a, Universidad Nacional Aut\'onoma de
M\'exico}

\begin{abstract}

We study the effect that the temperature structure has on the determination of
the primordial helium abundance, $Y_p$. We provide an equation linking
$T$(O~{\sc{iii}}), the temperature derived from the [O~{\sc{iii}}] lines, and
$T$(He~{\sc{ii}}), the temperature of the He I lines, both for H~{\sc{ii}}
regions with O$^{++}$ only and for H~{\sc{ii}} regions where a fraction of
O$^+$ is present. By means of $T$(He~{\sc{ii}}), which is always smaller than
$T$(O~{\sc{iii}}), we derive the helium abundances of 5 objects with low and
very low metallicity (NGC~346, NGC~2363, Haro~29, SBS~0335--052, and I~Zw~18);
these objects were selected from the literature because they include the 3 low
metallicity objects with the best line determinations and the 2 objects with
the lowest metallicity. {From} these abundances we obtain that $Y_p$(nHc) $ =
0.2356 \pm 0.0020$, a value 0.0088 lower than that derived by using
$T$(O~{\sc{iii}}).  We call this determination $Y_p$(nHc) because the
collisional contribution to the Balmer line intensities has not been taken into
account. All the recent $Y_p$ determinations in the literature have not taken
into account the collisional contribution to the Balmer line intensities. By
considering the collisional contribution to the Balmer line intensities of
these five objects we derive that $Y_p$(+Hc)$ = 0.2384 \pm 0.0025$.

\end{abstract}


\section{Introduction}

The determination of the pregalactic, or primordial, helium abundance by mass
$Y_p$ is paramount for the study of cosmology, the physics of elementary
particles, and the chemical evolution of galaxies \citep[e.g.] [and references
therein]{fie98,izo99,pei99y}. In this paper we present a new determination of
$Y_p$ based on observations of the metal-poor extragalactic H~{\sc{ii}} regions
NGC~346, I~Zw~18, NGC~2363, Haro~29, and SBS~0335--052. This determination is
compared with those carried out by other authors. A preliminary account on the
results for I~Zw~18 and NGC~2363 is presented elsewhere \citep*{pea01}.

There are three problems affecting the $Y_p$ determination that need to be
further analyzed: $a$) the temperature structure, $b$) the ionization
structure, and $c$) the collisional excitation of the hydrogen lines. This
paper will be mainly concerned with the effect of the temperature structure of
the nebulae on the helium line intensities. In a future paper we will discuss
in more detail the problem of the collisional excitation of the hydrogen lines
\citep*{lur01}. The most accurate $Y_p$ values in the literature have been
derived under the assumption of no contribution to the hydrogen Balmer lines
due to collisional excitation, we will refer to these determinations in this
paper as $Y_p$(nHc) and to those that include the hydrogen collisional effect
as $Y_p$(+Hc). Due to the collisional contribution to the Balmer line
intensities the $Y_p$(nHc) determinations are lower limits to the $Y_p$(+Hc)
value.

There are several pieces of evidence indicating that $T$(He~{\sc{ii}}) is
smaller than $T$(O~{\sc{iii}}): $a$) observations of NGC~346 \citep*
[hereinafter Paper~I] {pei00z}, $b$) photoionization models of giant
extragalactic H~{\sc{ii}} regions \citep{sta99, lur99, lur01, rel01}, $c$) in
photoionization models $T$(O~{\sc{ii}}) is smaller than $T$(O~{\sc{iii}}) for
objects with $T$(O~{\sc{iii}}) higher than 12,360~K \citep{sta90}, and a
considerable fraction of the He~{\sc{i}} lines originates in the O$^+$
zone. This small difference is significative because it lowers the He$^+$/H$^+$
determination. By combining the observed He~{\sc{i}} lines it is found that
$T$(He~{\sc{ii}}) and $N_e$(He~{\sc{ii}}) are coupled in the sense that a lower
$T$ implies a higher $N_e$ reducing the He$^+$/H$^+$ determination (see
Paper~I).

Another problem that has to be considered to determine very accurate He/H
values of a given H~{\sc{ii}} region is its ionization structure. The total
He/H value is given by:

\begin{eqnarray}
\frac{N ({\rm He})}{N ({\rm H})} & = &
\frac {\int{N_e N({\rm He}^0) dV} + \int{N_e N({\rm He}^+) dV} + 
\int{N_e N({\rm He}^{++})dV}}
{\int{N_e N({\rm H}^0) dV} + \int{N_e N({\rm H}^+) dV}},
						\nonumber \\
& = & ICF({\rm He})
\frac {\int{N_e N({\rm He}^+) dV} + \int{N_e N({\rm He}^{++}) dV}}
{\int{N_e N({\rm H}^+) dV}}
\label{eICF}
.\end{eqnarray}

For objects of low degree of ionization it is necessary to consider the
presence of He$^0$ inside the H$^+$ zone, while for objects of high degree of
ionization it is necessary to consider the possible presence of H$^0$ inside
the He$^+$ zone. For objects of low degree of ionization $ICF({\rm He})$ might
be larger than 1.00, while for objects of high degree of ionization $ICF({\rm
He})$ might be smaller than 1.00.  This problem has been discussed by many
authors
\citep[e.g.][]{shi74,sta83,pen86,vil88,pag92,arm99,pei00y,vie00,vig00,bal00y,
sau01}.

Each H~{\sc{ii}} region is different and a good photoionization model is needed
to estimate an $ICF({\rm He})$ of very high accuracy. In this paper we will
assume that the H~{\sc{ii}} regions are chemically homogeneous.

In section 2 we will adapt the $t^2$ and $T_0$ formalism introduced by
\citet{pei67} to relate $T$(O~{\sc{iii}}) and $T$(O~{\sc{ii}}) with
$T$(He~{\sc{ii}}), the full mathematical treatment is presented in the
Appendix. In section 3 we will rediscuss the $Y$ determination of NGC 346, the
most luminous H~{\sc{ii}} region in the SMC, carried out in Paper~I.  In
sections 4 to 7 we redetermine the He$^+$/H$^+$ values for I~Zw~18, NGC~2363,
Haro~29, and SBS~0335--052, based on the observations of \citet*{izo97} and
\citet{izo99} and on photoionization models for these objects. These four
objects were selected from their sample based on the following criteria: the
two objects with the lowest heavy element abundances and the two objects with
the smallest observational errors \citep{pei01}. We derive He$^+$/H$^+$,
$N_e$(He~{\sc{ii}}), $T$(He~{\sc{ii}}), and $\tau$(3889) self-consistently
based on all the observed He~{\sc{i}} line intensities, with the exception of
those strongly affected by underlying absorption.  By combining the
He$^+$/H$^+$ values of these four objects with that of NGC 346 we derive a new
$Y_p$(nHc) in section 8. The collisional contribution to the Balmer lines is
estimated in section 9. The discussion and conclusions are presented in
sections 10 and 11.

\section{Temperatures \label{stemp}}

\subsection{$T$(O~{\sc{iii}}) and $T$(He~{\sc{ii}}) for a pure O$^{++}$ nebula}

We first consider an H~{\sc{ii}} region where the He~{\sc{ii}}, the
O~{\sc{iii}}, and the H~{\sc{ii}} volumes coincide perfectly.  If we further
assume that the temperature is constant we have a very simple model which gives
us a first approximation to a real nebula.

If we now assume that the H~{\sc{ii}} region has an average temperature
given by
\begin{equation}
T_0=\frac{\int T N_e N_p \, dV}{\int N_e N_p  \, dV}
\label{eT0def}
,\end{equation}
and a mean square temperature variation given by:
\begin{equation}
t^2=\frac{\int (T-T_0)^2 N_e N_p \, dV}{T_0^2 \int N_e N_p  \, dV}
\label{et2def}
,\end{equation}
we find that the observed $T$(4363/5007), derived from the $I$(4363)/$I$(5007)
ratio, is given by \citep{pei67}:
\begin{equation}
T({\rm O~{\scriptstyle{III}}})=T_0
\left( 1 + \left[\frac{90800}{T_0}-3 \right] \frac{t^2}{2} \right)
\label{eTO3-T0}
,\end{equation}
and that the temperature associated to the He~{\sc{i}} recombination lines, 
that originate in the  He~{\sc{ii}} region, is (see Paper~I):
\begin{equation}
T({\rm He~{\scriptstyle{II}}}) = T_0
	\left(1-1.4t^2\right)
\label{eTHe2-T0}
.\end{equation}
This means that the temperature that should be used to measure
the helium abundance is given by:
\begin{equation}
T({\rm He~{\scriptstyle{II}}}) = T({\rm O~{\scriptstyle{III}}})
	\left(1-\left[\frac{90800}{T({\rm O~{\scriptstyle{III}}})}-0.2\right]
	\frac{t^2}{2}\right)
\label{eTHe2-TO3}
.\end{equation}

In Figure~\ref{fTO3-THe}, and based on equation~\ref{eTHe2-TO3}, we present
$T$(He~{\sc{ii}})/$T$(O~{\sc{iii}}) versus $t^2$ for different values of
$T$(O~{\sc{iii}}).

\subsection{$T$(He~{\sc{ii}}) for a nebula with O$^+$ and O$^{++}$}

We consider now another H~{\sc{ii}} region where the He~{\sc{ii}} volume
coincides with the H~{\sc{ii}} volume, but now oxygen is present both in the
form of O$^{++}$ and O$^+$, with the O$^+$ weight described by the fraction
$\alpha$:

\begin{equation} 
\alpha= \frac{\int N_e N({\rm O}^+) \, dV}
       {\int N_e N({\rm O}^+) \, dV \, + \, \int N_e N({\rm O}^{++}) \, dV}
\label{ealpha}
.\end{equation}

If the average temperatures in the O~{\sc{ii}} and O~{\sc{iii}} zones are given
by:
\begin{equation} T_{02}=\frac{\int T N_e N({\rm O}^+) \, dV}
{\int N_e N({\rm O}^+) \, dV}
\label{eT02def}
\end{equation}
and
\begin{equation} T_{03}=\frac{\int T N_e N({\rm O}^{++}) \, dV}
{\int N_e N({\rm O}^{++}) \, dV}
\label{eT03def}
\end{equation}
respectively, then the average temperature for the whole
ionized region  $T_0$ is equal to
\begin{equation}
T_0=\alpha T_{02}+\left(1-\alpha\right)T_{03}
\label{eT0-T02+T03}
;\end{equation}
the fractional difference between $T_{03}$ and $T_0$ can thus be characterized
by the parameter $\beta$ so that
\begin{equation}
\frac{T_{03}}{T_0}=1+\beta
\label{ebeta_1}
,\end{equation}
and the ratio between $T_{02}$ and $T_0$ is:
\begin{equation}
\frac{T_{02}}{T_0}
=1-\beta \left( \frac{1-\alpha}{\alpha} \right)
=1+\beta-\frac{\beta}{\alpha}
\label{ebeta_2}
.\end{equation}

In the general case when the O~{\sc{ii}} and O~{\sc{iii}} zones are not of
uniform temperature, but include temperature fluctuations, represented by:
\begin{equation}
t_2^2=\frac
{\int (T-T_{02})^2 N_e N({\rm O}^+) \, dV}
{T_{02}^2 \int N_e N({\rm O}^+) \, dV}
\label{et22def}
\end{equation}
and
\begin{equation}
t_3^2=\frac
{\int (T-T_{03})^2 N_e N({\rm O}^{++}) \, dV}
{T_{03}^2 \int N_e N({\rm O}^{++}) \, dV}
\label{et32def}
\end{equation}
respectively, the mean square temperature fluctuation for the entire
H~{\sc{ii}} region, $t^2$, is given by:
\begin{equation}
t^2	= 	  \alpha \left( \frac{T_{02}-T_0}{T_0}\right)^2
		+ \left(1-\alpha\right) \left( \frac{T_{03}-T_0}{T_0}\right)^2
		+ \alpha t_2^2 \left(\frac{T_{02}}{T_0}\right)^2
		+ \left(1-\alpha\right) t_3^2 \left(\frac{T_{03}}{T_0}\right)^2
\label{et2ini}
,\end{equation} were the first two terms represent the deviation from uniform
temperature given by the difference in temperature between each zone and the
average for the whole H~{\sc{ii}} region, while the other two represent the
additional contribution due to inhomogeneities in each zone.

Substituting the first 2 terms with equations \ref{ebeta_2} and \ref{ebeta_1}
and by keeping the last 2 terms to second order we obtain:
\begin{eqnarray}
t^2	& = 	& \alpha \beta^2 \left(\frac{1-\alpha}{\alpha}\right)^2
		+ \left(1-\alpha\right) \beta^2
		+ \alpha t_2^2
		+ \left(1-\alpha\right) t_3^2
                                                            \nonumber \\
	& = 	& \beta^2 \left(\frac{1-\alpha}{\alpha}\right)
		+ \alpha t_2^2
		+ \left(1-\alpha\right) t_3^2
\label{et2}
.\end{eqnarray} $T_{03}$ and $t_3^2$ are related to $T$(O~{\sc{iii}}) by (see
equation~\ref{eTO3-T0}):
\begin{equation}
T({\rm O~{\scriptstyle{III}}})=T_{03}
\left( 1 + \left[\frac{90800}{T_{03}}-3 \right] \frac{t_3^2}{2} \right)
\label{eTO3-T03}
;\end{equation}
while $T_{02}$ and $t_2^2$ are related to an O~{\sc{ii}} temperature,
determined from the $I$(3727)/$I$(7325) ratio, that
in the low density limit is given by \citep{pei67}:
\begin{equation}
T({\rm O~{\scriptstyle{II}}})=T_{02}
\left( 1 + \left[\frac{97300}{T_{02}}-3 \right] \frac{t_2^2}{2} \right)
\label{eTO2-T02}
.\end{equation}

It is convenient to obtain an equation, with the same physical meaning than
that of equations \ref{eTO3-T03} and \ref{eTO2-T02}, that relates $T_0$ and
$t^2$ with a temperature that can be derived from observations, particularly
with those of the O~{\sc{ii}} and O~{\sc{iii}} forbidden lines. Intuitively we
can define the following equation (the formal derivation of
equation~\ref{eTO2+3_T0m} is given in the Appendix):
\begin{equation}
T({\rm O~{\scriptstyle{II+III}}})=T_0
\left( 1 + \left[\frac{\theta}{T_0}-3 \right] \frac{t^2}{2} \right)
\label{eTO2+3_T0m}
,\end{equation}
where $T_0$ and $t^2$ correspond to the values for the entire
volume, $\theta$ represents an effective average over the excitation energy
needed to produce the oxygen lines,
\begin{equation}
\theta=\left(1-\alpha\right)90800+\alpha\,97300
\label{etheta}
\end{equation}
and $T({\rm O~{\scriptstyle{II+III}}})$ is an average between $T$(O~{\sc{ii}})
and $T$(O~{\sc{iii}}). However, it can be shown (see Appendix) that, the simple
average of the temperatures weighted by $\alpha$ and $1-\alpha$ does not give
an equation accurate to second order and that $T({\rm
O~{\scriptstyle{II+III}}})$ needs a factor $\Gamma$ of the form:
\begin{equation}
T({\rm O~{\scriptstyle{II+III}}})= \left[\left(1-\alpha\right)T({\rm
O~{\scriptstyle{III}}}) +\alpha \, T({\rm O~{\scriptstyle{II}}})\right]\Gamma
\label{eTO2+3def1}
,\end{equation}
$\Gamma$ is very close to unity for objects where $\alpha(1-\alpha)<0.08$
or where $T$(O~{\sc{ii}}) is similar to $T$(O~{\sc{ii}})
($0.9<T({\rm O~{\scriptstyle{II}}})/T({\rm O~{\scriptstyle{III}}})<1.1$); but
for H~{\sc{ii}} regions where there are appreciable fractions of both O$^+$ and
O$^{++}$ and where there is an appreciable difference between $T$(O~{\sc{iii}})
and $T$(O~{\sc{ii}}), $\Gamma$ must be calculated as described in the Appendix.
(All computations done in this paper were done using the formalism described in
the Appendix.)

Finally, solving equation~\ref{eTO2+3_T0m} for $T_0$ and substituting it in
equation \ref{eTHe2-T0} it follows that
\begin{equation}
T({\rm He~{\scriptstyle{II}}}) = T({\rm O~{\scriptstyle{II+III}}})
		\left(1-\left[\frac{\theta}
		{T({\rm O~{\scriptstyle{II+III}}})}-0.2\right]
		\frac{t^2}{2}\right)
\label{eTHe2-TO2+3}
,\end{equation}
which can now be used to determine the He abundances. To derive $T({\rm
O~{\scriptstyle{II+III}}})$ from the observed $T$(O~{\sc{ii}}) and
$T$(O~{\sc{iii}}) values we need to use equations \ref{ealpha}, \ref{etheta},
and \ref{eTO2+3def1} for $\Gamma=1$; for $\Gamma$ larger than one we also need
to use equations \ref{eGamma}, \ref{edelta_1}, and~\ref{egamma}. If only
$T$(O~{\sc{iii}}) is available we need a photoionization model to obtain a
relationship between $T$(He~{\sc{ii}}) and $T$(O~{\sc{iii}}).

In Figure~\ref{fTO2+3} we present the relationship between $T$(He~{\sc{ii}})
and $T$(O~{\sc{iii}}) as a function of $t^2$ for objects with different O$^+$
fractions. Figure~\ref{fTO2+3} is based on equations \ref{ealpha},
\ref{etheta}, \ref{eTO2+3def1}, \ref{eGamma}, \ref{edelta_1}, and~\ref{egamma}
and the relationship
\begin{equation}
T_{02} = 2430 + T_{03}(1.031 - T_{03}/54350)
\label{eStasinska}
,\end{equation}
derived from the photoionization models by \citet{sta90}. {From} this equation
it is obtained that: for $T_{03}$ = 12,360~K then $T_{03} = T_{02}$, for higher
temperatures $T_{03} > T_{02}$, and for lower temperatures $T_{03} < T_{02}$.

{From} Figure~\ref{fTO2+3} and equation~\ref{eStasinska} we note the following
cases for objects with $T$(O~{\sc{iii}}) $>$ 12,360~K: $a$) For $\alpha = 0.00$
and $t_{3}{^2} = 0.000$, then $T$(He~{\sc{ii}}) = $T$(O~{\sc{iii}}) and $t^2 =
0.000$.  $b$) For $\alpha$ $\neq 0.00$ and $t_{2}{^2} = t_{3}{^2} = 0.000$,
then $T$(He~{\sc{ii}}) is {\it always} smaller than $T$(O~{\sc{iii}}) and $t^2$
is {\it always} larger than 0.000.  $c$) For $\alpha = 0.00$, and $t_{3}{^2} >
0.000$, then $T$(He~{\sc{ii}}) $<$ $T_e$(O~{\sc{iii}}) and $t^2 > 0.000$.  $d$)
For $\alpha$ $\neq 0.00$ and $t_{2}{^2} \sim t_{3}{^2} > 0.000$, then
$T$(He~{\sc{ii}}) is even smaller than in case $b$ and $t^2$ is even larger
than in case $b$.

Very metal-poor extragalactic H~{\sc{ii}} regions, those that have been used to
derive the primordial helium abundance, have $T$(O~{\sc{iii}}) larger than
12,360~K and are best represented by case $d$.

Figure~\ref{fTO2+3} is also related to the Averaged Quantities block of {\sc
Cloudy}'s output \citep{fer96,fer98,kin95}.  In that section of {\sc Cloudy}'s
output there is a line labeled "Peimbert" with the following entries:
"T(O~IIIr)" that corresponds to $T$(O~{\sc{iii}}) in this paper; "T(Bac)" is
the hydrogen temperature resulting from the predicted Balmer jump and H$\beta$
for the optically thick case, corresponds to the outer surface of the object
and does not represent typical H~{\sc{ii}} regions; "T(Hth)" is the hydrogen
temperature resulting from the predicted Balmer jump and H$\beta$ for the
optically thin case, corresponds to $T$(Bac) defined by \citet{pei67} and used
by many authors, and applies to typical H~{\sc{ii}} regions; "t$^2$(Hstr)" is
the structural value computed across the H~{\sc{ii}} region, and it is called
$t^2$ in this paper.  "T(O3-BAC)" and "t$^2$(O3-BC)" are the average
temperature and the mean square temperature fluctuation values derived from the
predicted $T$(O~{\sc{iii}}) and $T$(Bac) values, they correspond to $T_0$ and
t$^2$(Hstr) only when all the O is in the O$^{++}$ stage. In the presence of
O$^+$ and according to Figure~\ref{fTO2+3} and equation~\ref{eStasinska} we can
distinguish two cases: when $T$(O~{\sc{iii}}) is higher than 12,360~K then
"t$^2$(O3-BC)" is higher than $t^2$, but when $T$(O~{\sc{iii}}) is smaller than
12,360~K then "t$^2$(O3-BC)" is smaller than $t^2$.

{From} photoionization models of H~{\sc{ii}} regions it has been found that
$t^2$ is in the 0.002 to 0.03 range, with typical values around 0.005
\citep[e.g.][]{gru92,kin95,per97}; from observations it has been found that
$t^2$ is in the 0.01 to 0.04 range, with typical values around 0.03
\citep[e.g.][and references therein]{pei95,pei99z}.  As a first approximation
due to equation \ref{eStasinska} we expect minimum $t^2$ values from
photoionized models with $T\sim$ 12,360~K and an increase in $t^2$ for hotter
and cooler models. Most of the observational determinations of $t^2$ come from
H~{\sc{ii}} regions with $T$(O~{\sc{iii}}) $\sim$ 8000~K which are biased
toward one of the higher ends of the $t^2$ distribution. Very metal-poor
H~{\sc{ii}} regions are biased to the other high end of the $t^2$ distribution,
particularly those with WR stars and a large fraction of O$^+$.  The $t^2$
values derived from photoionization models correspond to $T$(He~{\sc{ii}})
values 3\% to 11\% smaller than $T$(O~{\sc{iii}}). In the presence of
additional sources of energy, to those provided by photoionization, the
difference between $T$(He~{\sc{ii}}) and $T$(O~{\sc{iii}}) might be larger.

\section{NGC 346}

In Paper~I we have analyzed NGC~346, the most luminous H~{\sc{ii}} region in
the Small Magellanic Cloud. Due to the relatively small distance to NGC~346 it
is possible to place the observing slit avoiding the brightest stars, thus
minimizing the effect of the underlying absorption in the He$^+$/H$^+$
determination.

The determination of $Y_p$ based on NGC~346 can have at least four significant
advantages and one disadvantage with respect to those based on distant
H~{\sc{ii}} region complexes: $a$) no underlying absorption correction for the
helium lines is needed because the ionizing stars can be excluded from the
observing slit, $b$) the determination of the helium ionization correction
factor can be estimated by observing different lines of sight, $c$) the
accuracy of the determination can be estimated by comparing the results derived
from different lines of sight, $d$) the electron temperature is generally
smaller than those of metal poorer H~{\sc{ii}} regions reducing the effect of
collisional excitation from the metastable 2$^3$ S level of He~{\sc{i}} and
from the ground level of H~{\sc{i}}, and $e$) the disadvantage is that the
correction due to the chemical evolution of the SMC is in general larger than
for the other systems.

{From} the observations of the He~{\sc{i}} lines $\lambda\lambda$ 3889, 4026,
4387, 4471, 4922, 5876, 6678, 7065, and 7281 (see Table~\ref{tlines}) and from
the temperatures associated with the O~{\sc{iii}} lines and the Balmer
discontinuity in region A we have used a maximum likelihood method to determine
simultaneously and self-consistently several parameters. Our results are:
$N_e$(He~{\sc{ii}})$ = 144^{+44}_{-38}\; {\rm cm}^{-3}$, $T$(He~{\sc{ii}})$ =
11,860 \pm 370\; {\rm K}$, ${\rm He/H}=0.07910 \pm 0.00072$, and $\tau$(3889) =
0.0. For this temperature range we obtain that O/H$ = (141\pm 21) \times
10^{-6}$. In Table~\ref{tchi} we present ${\rm He/H}$ and $\chi^2$ as a
function of several $N_e$(He~{\sc{ii}}) and $T$(He~{\sc{ii}}) values, which
include a few representative temperatures and the densities favored by those
temperatures. {From} this table it can be seen that the minimum $\chi^2$ values
imply a strong correlation between the temperature and the density, in the
sense that the lower the temperature, the higher the density, and consequently
the lower the ${\rm He/H}$ value.  In Paper~I it is also found that
$T_e$(He~{\sc{ii}}) is $8.6\% \pm 3\%$ smaller than $T_e$(O~{\sc{iii}}).  The
parameters that represent this model are presented in Table~\ref{tparameters}.

In Table~\ref{t346} we present the $Y$(nHc) values for different temperatures
and densities, where we have included the He$^{++}$/H$^+$ ratio (see Paper~I)
and an $ICF({\rm He}) = 1.00$.

\citet{rel01} have computed photoionization models for NGC~346 with {\sc
Cloudy}, they have found: $a$) that there is a negligible amount of H$^0$
inside the He$^+$ zone, in agreement with Paper~I, $b$) that $T$(He~{\sc{ii}})
is 5\% smaller than $T$(O~{\sc{iii}}) in their best model, and $c$) that the
$T$(O~{\sc{iii}}) computed by the models is 9\% smaller than the observed
value, indicating the presence of additional heating sources not included in
the photoionization models and probably implying that the real difference
between $T$(He~{\sc{ii}}) and $T$(O~{\sc{iii}}) is higher than predicted by the
models.

\section{I~Zw~18}

Up to now, with the exception of NGC~346, it has not been possible to derive
the $ICF({\rm He})$, $N_e$(He~{\sc{ii}}), $T$(He~{\sc{ii}}), $\tau$(3889), and
the He$^+$/H$^+$ ratio based only on the helium lines. Usually the observed
$T$(O~{\sc{iii}}) value has been used to complement the information provided by
the He~{\sc{i}} lines, moreover it has also been assumed that the observed
$T$(O~{\sc{iii}}) is equal to $T$(He~{\sc{ii}}). For I~Zw~18, SBS~0335--052,
NGC~2363, and Haro~29 we will make use of the observed $T$(O~{\sc{iii}}), but
we will not assume that it is equal to $T$(He~{\sc{ii}}), instead we will make
use of photoionization models computed with {\sc Cloudy} to relate both
quantities.

In general photoionization models predict $T$(O~{\sc{iii}}) values smaller than
observed \citep{sta99,lur99,lup01,lur01,rel01} indicating the possible presence
of an additional heating source not considered by the models; this result also
implies that the $T$(He~{\sc{ii}}) values predicted by the models will be
smaller than those $T$(He~{\sc{ii}}) values derived from the observed
$T$(O~{\sc{iii}}) values. Since the lower the temperature, the lower the
derived He/H value (see for example Table~\ref{tZw}), the use of the
temperatures predicted by the models yields spuriously low He/H values. On the
other hand the models can give us an estimate of $t^2$(O~{\sc{iii}}), probably
a lower limit. The formalism presented in section~\ref{stemp}, together with
$t^2$(O~{\sc{iii}}), provides a relationship between the observed
$T$(O~{\sc{iii}}) and $T$(He~{\sc{ii}}).

{From} photoionization models of I~Zw~18 based on {\sc Cloudy} \citep{lur01} we
find that: $ICF({\rm He}) = 1.00$, $\tau(3889) = 0.010 \pm 0.005$, and
$t^2$(O~{\sc{iii}})$= 0.013\pm 0.006$.  We consider $t^2$(O~{\sc{iii}}) as a
lower limit to $t^2$ (see equations \ref{et2}~and~\ref{et22t32}).  {From} the
formalism of section~\ref{stemp} together with $\alpha=0.265$ and
$T$(O~{\sc{iii}}) $= 19,060$~K \citep{izo99} we obtain $t^2= 0.024\pm 0.006$
and consequently that $T$(He~{\sc{ii}}) is 10.2\% smaller than
$T$(O~{\sc{iii}}).

To derive the He$^+$/H$^+$ ratio we have used the maximum likelihood method,
MLM (see Paper~I). The inputs are: $a$) $\tau(3889) = 0.010 \pm 0.005$, $b$)
$t^2= 0.024\pm 0.006$, $c$) $T$(O~{\sc{iii}})$= 19,060$~K and $d$) the
observations of $\lambda\lambda$ 3889, 4026, 4471, 5876, 6678, and 7065
by \citet{izo99}, where we have adopted the line intensity ratios presented in
Table~\ref{tlines}, corrected by underlying absorptions of 3.3\AA, 0.8\AA, and
0.8\AA\, for $\lambda\lambda$ 3889, 4026 and 4471 respectively, that correspond
to a burst model of 4~Myr \citep*{gon99}. The underlying absorptions predicted
by the burst model have been reduced by about 10\% to take into account the
presence of the nebular continuum emission; a similar reduction has been
applied to the underlying absorptions predicted by the starburst models for
NGC~2363, Haro~29, and SBS~0335--052.

To obtain He$^+$/H$^+$ values we need a set of effective recombination
coefficients for the He and H lines, the contribution due to collisional
excitation to the helium line intensities, and an estimate of the optical depth
effects for the helium lines. The recombination coefficients that we used were
those by \citet{sto95} for H, and \citet{smi96} for He. The He~{\sc{i}}
collisional contribution was estimated from \citet{kin95} and
\citet*{ben99}. The optical depth effects in the He~{\sc{i}} triplet lines were
estimated from the computations by \citet{rob68}, and based on Paper~I we will
assume that the He~{\sc{i}} singlet lines are produced under case B.

The maximum likelihood solution amounts to $a$) $N_e$(He~{\sc{ii}})$ = 87
^{+94}_{-78}\; {\rm cm}^{-3}$, $b$) He$^+$/H$^+ = 0.07673 \pm 0.00312$, $c$)
$\chi^2 = 4.2$, and $d$) $T_e$(He~{\sc{ii}})$= 17,120 \pm 600\; {\rm K}$,
10.2\% smaller than $T$(O~{\sc{iii}}) (see Table~\ref{tparameters}). For this
temperature range we obtain that O/H $= (19 \pm 2) \times 10^{-6}$. There is
not enough information to independently derive the $t^2$ and $\tau$(3889)
values from the observed lines in this object, therefore their output values
are equal to the input values. Table~\ref{tZw} presents the $Y$(nHc) values for
four $t^2$ values: 0.024 (the optimum value), 0.030 (+~1$\sigma$), 0.018
(-~1$\sigma$), and 0.006 (the minimum possible value given by the size and
temperature of the O$^+$ zone); and 6 different densities that include those
provided by the MLM for each $t^2$. The $Y$(nHc) values include the
contribution due to He$^{++}$ and an $ICF({\rm He}) = 1.00$.

\section{NGC~2363}

The $T$(O~{\sc{iii}}) values for the four objects observed by \citet{izo97} and
\citet{izo99} were derived by us from the observed line intensities, we obtain
the same results for I~Zw~18 and SBS~0335--052 than \citet {izo99}, but we
obtain values 650~K higher for NGC~2363 and Haro~29 than \citet{izo97}.

\citet{lur99} produced detailed photoionization models of NGC~2363, they find
also that the $T$(O~{\sc{iii}}) predicted by the models is considerably smaller
than 15,750~K, the observed value; from their models they find also that $t^2$
is in the 0.007 to 0.029 range. This amounts to a $T$(He~{\sc{ii}}) 6\% $\pm$
2\% smaller than $T$(O~{\sc{iii}}). {From} the models of NGC~2363 and the slit
used by \citet{izo97} we also find that: $ICF({\rm He}) = 0.993$, indicating
the presence of neutral hydrogen inside the ionized helium region (see
equation~\ref{eICF}), and $\tau(3889) = 0.50 \pm 0.15$.

Support for the relative high density comes from the $N_e$[S~{\sc{ii}}] derived
by \citet{izo97} and by \citet{est01} that amount to $120\; {\rm cm}^{-3}$ and
$340 \pm 120\; {\rm cm}^{-3}$ respectively and the $N_e$ [Ar~{\sc{iv}}] values
in the 400 to 800~cm$^{-3}$ range derived by \citet*{per01} and the $1300
\pm 500\; {\rm cm}^{-3}$ value derived by \citet{est01}; the [S~{\sc{ii}}]
lines trace the outer parts while the [Ar~{\sc{iv}}] lines trace the inner
parts of the H~{\sc{ii}} region.

For the MLM we use the following inputs: $a$) $\tau(3889) = 0.5 \pm 0.3$, $b$)
$t^2= 0.018\pm 0.011$, $c$) $T$(O~{\sc{iii}})$= 15,750$~K, and $d$) the
observations of $\lambda\lambda$ 3820, 3889, 4026, 4387, 4471, 5876, 6678,
7065, and 7281 by \citet{izo97} slightly modified to take into account
underlying absorptions of 0.3\AA, 1.8\AA, 0.4\AA, 0.2\AA, and 0.4\AA\, for
$\lambda\lambda$ 3820, 3889, 4026, 4387, and 4471 respectively (see
Table~\ref{tlines}), that correspond to a burst model of 3~Myr
\citep{gon99}. We did not consider $\lambda$4922 because its intensity was many
$\sigma$ away from the solution increasing $\chi^2$ to unacceptable values; we
also adopted an error for $\lambda$3889 three times larger than the one given
by \citet{izo97} due to our estimate of the combined errors produced by the
reddening correction and the underlying absorption correction.

With these inputs we obtain $a$) $N_e$(He~{\sc{ii}})$ = 250 ^{+88}_{-77}\; {\rm
cm}^{-3}$, $b$) $t^2= 0.021\pm 0.010$, $c$) $\tau(3889) = 0.98 \pm 0.25$, $d$)
He$^+$/H$^+ = 0.07997 \pm 0.00153$, $e$) $\chi^2 = 12.9$, and $f$)
$T$(He~{\sc{ii}})\, $= 14,710 \pm 440\; {\rm K}$, a value 6.5\% smaller than
$T$(O~{\sc{iii}}) (see Table~\ref{tparameters}).

Note that the $t^2$ and $\tau(3889)$ output values are not the same as the
input values. This is because the quality of the measurements for NGC~2363 is
better than that of I~Zw~18, and thus the values of $t^2$ and $\tau(3889)$ can
be directly estimated from the lines, however these values are not as accurate
as those of NGC~346 and the error bars would be very large, therefore the
observed line intensities get mixed with the $t^2$ and $\tau(3889)$ input
values in the MLM, giving us a different output.

For this temperature distribution we obtain O/H\,$= (95 \pm 8) \times
10^{-6}$. Table~\ref{t2363} presents the $Y$(nHc) values for four $t^2$ values:
0.021 (the optimum value), 0.031 (+~1$\sigma$), 0.011 (-~1$\sigma$), and
0.001 (the minimum possible value given by the size and temperature of the
O$^+$ zone); and 6 different densities that include those provided by the MLM
for each $t^2$. The $Y$(nHc) values include the contribution due to He$^{++}$
and an $ICF({\rm He}) = 0.993$.

\section{Haro~29}

Haro~29, also known as I~ZW~36 or Markarian~209, is one of the two blue compact
galaxies with the smallest observational errors in the list of \citet{izo97},
the other being NGC~2363.

{From} photoionization models of Haro~29 based on {\sc Cloudy} by \citet{lur01}
and the slit used by \citet{izo97} it is again found that the $T$(O~{\sc{iii}})
values predicted by the models are in the 14,550 to 15,100~K range, values
considerably smaller than the observed value of 16,050~K; from these models it
is also found that $ICF({\rm He}) =$ 0.995~- 0.996, $\tau(3889) \approx 1.4$,
and $t^2=$ 0.002~-0.004.

For the MLM we use the following inputs: $a$) $\tau(3889) = 1.43 \pm 0.715$,
$b$) $t^2= 0.020\pm 0.007$, and $c$) the observations of $\lambda\lambda$ 3820,
3889, 4026, 4387, 4471, 5876, 6678, 7065, and 7281 by \citet{izo97} slightly
modified to take into account underlying absorptions of 0.4\AA, 2.2\AA, 0.6\AA,
0.3\AA, and 0.6\AA\, for $\lambda\lambda$ 3820, 3889, 4026, 4387, and 4471
respectively (see Table~\ref{tlines}), that correspond to a burst model of
3.2~Myr \citep{gon99}.  We did not consider $\lambda$ 4922 because its
intensity was many $\sigma$ away from the solution increasing $\chi^2$ to
unacceptable values; we also adopted errors for $\lambda\lambda$3889 and 7065
three and two times larger than the one given by \citet{izo97} due to our
estimate of the combined errors produced by the reddening correction and the
underlying absorption correction.

{From} these inputs we obtain: $N_e$(He{~\sc{ii}})$ = 45^{+77}_{-66}\; {\rm
cm}^{-3}$, $\tau(3889) = 0.899 \pm 0.296$, $t^2= 0.019\pm 0.007$, and
He$^+$/H$^+ = 0.08157 \pm 0.00188$. The density result is of great concern, not
only is the lower limit unphysical, but the derived value is similar to the
root mean square density, $N_e(rms)$, that amounts to $=47\pm 7\; {\rm
cm}^{-3}$.  Giant H~{\sc{ii}} regions usually show $N_{\rm e}(rms)$ values
considerably smaller than those densities determined through a forbidden-line
ratio or from the He~{\sc{i}} lines based on the maximum likelihood method,
$N_{\rm e}^2(local)$; to quantify the difference the filling factor $\epsilon$
is used and is defined by means of the relation:
\begin{equation}
    N_{\rm e}^2(rms)=\epsilon N_{\rm e}^2(local)
\label{eepsilon}
.\end{equation}
Typical values of $\epsilon$ for giant H~{\sc{ii}} regions are in the 0.001 to
0.1 range \citep[e.g. Paper~I;][]{lur99,lup01,rel01} that would imply a
$N_e$(He~{\sc{ii}}) in the 149 to 1490~cm$^{-3}$ range. Since the density
derived from the MLM implies an $\epsilon \approx 1.00$, an unacceptable value,
we decided to make another estimate of $N_e$(He~{\sc{ii}}).

{From} the {\sc Cloudy} models by \citet{lur01} we found that a good fit to the
observed line intensities is obtained with $\epsilon \approx 0.04$, which
corresponds to an $N_e$(He~{\sc{ii}})$ = 235 \pm 85\; {\rm cm}^{-3}$. Fixing
this density as our preferred value we obtain $\tau(3889) = 0.41 \pm 0.30$,
$t^2= 0.023\pm 0.007$ (see Table~\ref{tparameters}), and He$^+$/H$^+ = 0.07800
\pm 0.00178$, other values can be obtained or interpolated from
Table~\ref{th29}. The $Y$(nHc) values include the contribution due to He$^{++}$
and an $ICF({\rm He}) = 0.995$.

For this temperature distribution we obtain O/H\,$= (78 \pm 10) \times
10^{-6}$.

\section{SBS~0335--052}

After I~Zw~18, SBS~0335--052 is the extragalactic H~{\sc{ii}} region with the
second lowest metallicity known; and being brighter than I~Zw~18 makes it
critical in determining $Y_p$.

We have computed a series of {\sc Cloudy} models for SBS~0335--052
\citep{lur01}. We find that the $T$(O~{\sc{iii}}) predicted by the models is
$T$(O~{\sc{iii}})~= 17,500~- 18,500~K, again considerably smaller than the
observed value of 20,500~K.  For the three central values of the observing slit
used by \citet{izo99} (center, 0."6SW, 0."6NE) we have extracted from the
models the following predictions: $ICF({\rm He}) = 0.9985$, $\tau(3889) \approx
1.26$, and $t^2$ in the 0.010 to 0.015 range.

To correct for underlying absorption the blue helium lines and the Balmer lines
observed by \citet{izo99} we adopted the values predicted by \citet{gon99} for
a 3~Myr old instantaneous burst with $Z=0.001$. The corrections for
$\lambda\lambda$ 4026, 4471, 4922, 3889 and H$\beta$ amount to 0.4\AA, 0.4\AA
\, 0.3\AA \, 1.8\AA\, and 2.0\AA\ respectively. Note that \citet{izo99}
corrected the Balmer lines for an underlying absorption of 0.2\AA\, and that
they did not correct the helium lines. We adopted the $C$(H$\beta$) value
determined by \citet{izo99} that is mainly based on the
$I$(H$\alpha$)/$I$(H$\beta$) ratio, this ratio is not affected by underlying
absorption (but it might be affected by collisional excitation of the Balmer
lines, see below). The line ratios are presented in Table~\ref{tlines}.

{From} the MLM and the following inputs: $a$ $\tau(3889) = 1.26 \pm 0.63$, $b$)
$t^2= 0.020\pm 0.007$, and $c$) the modified observations of $\lambda\lambda$
3889, 4026, 4471, 5876, 6678, and 7065 ($\lambda$ 4922 was not considered
because its intensity was many $\sigma$ away from the solution increasing the
$\chi^2$ to unacceptable values); we obtain $N_e$(He~{\sc{ii}})$ =
297^{+65}_{-56}\; {\rm cm}^{-3}$, He$^+$/H$^+ = 0.07640 \pm 0.00152$,
$\tau(3889) = 1.60 \pm 0.35$, and $t^2= 0.021\pm 0.007$, this $t^2$ corresponds
to a $T$(He~{\sc{ii}})\, $= 19,290 \pm 310\; {\rm K}$ (see
Table~\ref{tparameters}).

For this temperature distribution we obtain O/H\,$= (24 \pm 3) \times
10^{-6}$. Table~\ref{t0335} presents the $Y$(nHc) values for four $t^2$ values:
0.021 (the optimum value), 0.028 (+~1$\sigma$), 0.014 (-~1$\sigma$), and
0.004 (the minimum possible value given by the size and temperature of the
O$^+$ zone); and 6 different densities that include those provided by the MLM
for each $t^2$. The $Y$(nHc) values include the contribution due to He$^{++}$
and an $ICF({\rm He}) = 0.9985$.

\section{Determination of $Y_p$(nHc)}

In Table~\ref{tcomp} we compare the $Y$(nHc) maximum likelihood values computed
in this paper for $t^2 \neq 0.000$ (see Tables~\ref{t346}-\ref{t0335}) with the
$Y$(nHc) values for $t^2 = 0.000$ computed by \citet{izo97,izo99}. The
differences in the $Y$ values between both types of determinations are of the
order of 0.0070, and are manly due to the treatment of the temperature. The
errors that we quote are larger than those by Izotov and colaborators, both
groups include the errors in the observed line intensity ratios, however we
also include the uncertanties on the $N_e$(He~{\sc{ii}}), $t^2$, and
$\tau(3889)$ values while they do not. None of the $Y$(nHc) errors include the
uncertanties due to the collisional excitation of the Balmer lines (which will
be discussed in the next section) nor the prescence of other possible
systematic effects. In Figure~\ref{fY-OH} we also present the $Y$(nHc) versus
O/H diagram for all the objects.

To determine the $Y_p$(nHc) value for all the objects it is necessary to
estimate the fraction of helium present in the interstellar medium produced by
galactic chemical evolution. We will assume that
\begin{equation}
Y_p  =  Y - O \frac{\Delta Y}{\Delta O}
\label{eDeltaO}
,\end{equation}
where $O$ is the oxygen abundance by mass. The $\Delta O$ baseline given by the
five objects in the sample is very small and consequently produces large errors
in the $\Delta Y/\Delta O$ determination, therefore we decided to adopt $\Delta
Y/\Delta O = 3.5 \pm 0.9$, the slope derived in Paper~I. {From} this slope and
the five $Y$(nHc) values for $t^2 \neq 0.000$ presented in Table~\ref{tcomp} we
derive $Y_p$(nHc)$ = 0.2356 \pm 0.0020$. In Table~\ref{tcomp} we present the
$Y_p$(nHc) determination derived from the $Y$(nHc) values for $t^2 = 0.000$ 
obtained by \citet{izo98} and \citet{izo99} for Haro~29, NGC~2363,  
I~Zw~18, and
SBS~0335--052,  and find that our result
is 0.0088 smaller than theirs. The main difference is due to our use of
$T$(He~{\sc{ii}}) instead of $T$(O~{\sc{iii}}) used by them.

\section{Collisional excitation of the hydrogen lines}

\citet{dav85} were the first to estimate the collisional contribution to the
Balmer lines and its effect on the determination of $Y_p$; they made a crude
estimate of this effect for I~Zw~18 and concluded that the contribution to
$I$(H$\alpha$) {\it may} be roughly 2\%. All the subsequent determinations of
$Y_p$ in the literature have been derived under the assumption of no
contribution to the hydrogen Balmer lines due to collisional excitation, we
have referred to these determinations in this paper as $Y_p$(nHc) and as
$Y_p$(+Hc) to those primordial helium abundance determinations that include the
collisional contribution effect. Further discussion of the relevance of the
collisional contribution to the Balmer lines is presented elsewhere
\citep{sta01,lur01}.

{From} observations it is not easy to estimate this effect for two reasons:
small changes are expected in the Balmer line ratios, and the increase in
$I$(H$\alpha$)/$I$(H$\beta$) due to collisions can be ascribed to a spuriously
higher $C$(H$\beta$). The He/H line ratios are affected by the increase of the
Balmer line intensities due to collisions and by the decrease of $C$(H$\beta$)
due to the increase of $I$(H$\alpha$)/$I$(H$\beta$).

{From} a series of {\sc Cloudy} models it is found that the collisional
contribution to $I$(H$\beta$) for I~Zw~18 and SBS~0335--052 is in the 2\% to
6\% range, for Haro~29 and NGC~2363 in the 1\% to 3\% range and for NGC~346 in
the 0.6\% to 1.2\% range. The effect of collisions on the Balmer line
intensities together with the {\sc Cloudy} models for I~Zw~18, SBS~0335--052,
and Haro~29 will be discussed extensively elsewhere \citep{lur01}. The {\sc
Cloudy} models for NGC~2363 and NGC~346 are those by \citet{lur99} and
\citet{rel01} respectively.

To estimate $Y$(+Hc) for NGC~346, NGC~2363, Haro~29, and I~Zw~18 we used the
published Balmer line intensities without modifying the $C$(H$\beta$)
determination, the reason is that the published $C$(H$\beta$) values are
already very low and do not seem to indicate that they have been overestimated.
For NGC~346 the published $C$(H$\beta$) value is 0.15$\pm 0.01$ and the value
derived from the embedded stellar cluster amounts to $0.19 \pm 0.01$ (see
Paper~I), for NGC 2363 and Haro~29 $C$(H$\beta$) amounts to 0.11 and 0.00
respectively \citep{izo97}, and for I~Zw~18 it amounts to $0.015 \pm 0.020$
\citep{izo99}.  The $Y$(+Hc) estimates are presented in Table~\ref{tcomp},
where the collisional contribution was obtained from the models mentioned in
the previous paragraph. The errors in the $Y$(+Hc) determinations include the
same errors as the $Y$(nHc) determinations plus the error produced by the
uncertainty in the collisional contribution effect, but they do not include the
errors due to other possible systematic effects.

Alternatively \citet{izo99} derived for the three central positions of
SBS~0335--052 a $C$(H$\beta$) value of $0.237 \pm 0.019$, but the value derived
from $I$(H$\alpha$)/$I$(H$\beta$) is higher than those derived from
$I$(H$\gamma$)/$I$(H$\beta$) and $I$(H$\delta$)/$I$(H$\beta$) (after correcting
the Balmer lines for underlying absorption as mentioned in section 7); this
difference could be due to collisional excitation of the Balmer lines,
therefore from the corrected Balmer lines for underlying absorption and from
the effect of collisions, as predicted by one of our {\sc Cloudy} models, we
obtained a $C$(H$\beta$) of 0.15; for this value of the reddening we corrected
all the line intensities and from them we derived the $Y$(+Hc) value presented
in Table~\ref{tcomp}.

Our preliminary result of $Y_p$(+Hc), is about 0.0028 larger than $Y_p$(nHc)
(see Table~\ref{tcomp}). {From} Table~\ref{tcomp} it can be seen that the
$Y$(+Hc) values for NGC~346, NGC~2363 and Haro~29 are similar to those for
I~Zw~18 and SBS~0335--052.  This result {\it may} indicate that the corrections
adopted for collisional excitation of the Balmer lines have been overestimated
because the production of helium due to galactic chemical evolution is expected
to be higher for the first three objects than for the last two (see
equation~\ref{eDeltaO}).

\section{Discussion}

The brightest extragalactic H~{\sc{ii}} regions with temperatures in the
14,000 to 16,000~K range might be the best objects to determine $Y_p$. The
reasons are the following: $a$) there are many of these objects available and
the brightest of them are brighter than the metal-poorest objects known which
implies that the errors in the line intensity ratios will be smaller, and $b$)
the effect of collisional excitation of the Balmer lines is considerably
smaller than for the metal-poorer objects (those with temperatures in the
18,000 to 22,000~K range) reducing the error in the $Y$ determination due to
this effect.

On the other hand the correction to the $Y$ determination to obtain $Y_p$ based
on objects with temperatures in the 14,000 to 16,000~K range increases the
error in the $Y_p$ determination, but in general the size of this error is
expected to be smaller than those introduced in the previous paragraph because
the correction is small and can be obtained with high accuracy under the
assumption that $\Delta Y/\Delta O = 3.5 \pm 0.9$ (see Paper~I). {From}
chemical evolution models it is found that changes in $\Delta Y/\Delta O$ are
small even in the presence of a very strong burst of star formation
\citep*{car99}; moreover these objects seem to have been forming stars over
periods of at least 1 to 2~Gyr implying that most of their oxygen formed before
the present burst, as in the case of Haro~29 \citep[e.g.][]{sch01}, and
consequently that the expected changes in $\Delta Y/\Delta O$ due to bursts are
very small.

The $Y_p$(nHc) value derived by us is significantly smaller than the value
derived by \citet{izo98}, from the $Y$ -- O/H linear regression for a sample of
45 BCGs, and by \citet{izo99}, from the average for the two most metal
deficient galaxies known (I~Zw~18 and SBS~0335--052), that amount to $0.2443
\pm 0.0015$ and $0.2452 \pm 0.0015$ respectively.

Our $T_e$(He~{\sc{ii}}) determinations, that we prefer, produce $Y$ values that
are about 0.007 smaller than those derived by \citet{izo98} and \citet{izo99}, 
a small but significative difference, as we will see below. It should be noted
that the abundances for NGC~2363, SBS~0335--052, Haro~29, and I~Zw~18 obtained
by both groups are based on the same observations.

{From} photoionization models computed with {\sc Cloudy} we estimate that
$T$(He~{\sc{ii}}), the temperature that should be used to determine the helium
abundance, should be at least 5\% smaller than $T$(O~{\sc{iii}}). Moreover, if
in addition to photoionization there is additional energy injected to the
H~{\sc{ii}} region the difference between $T$(O~{\sc{iii}}) and
$T$(He~{\sc{ii}}) might be even larger.

Figure~\ref{feta} shows the helium, deuterium, and lithium abundances predicted
by standard Big Bang nucleosynthesis computations with three light neutrino
species for different values of $\eta$, the baryon to photon ratio
\citep{tho94,fio98}, also in this figure we present observational abundance
determinations of these elements. The implications of this figure for the
determination of $\Omega_b$, the baryon content of the Universe, are presented
in Table~\ref{tbar}.

\section{Conclusions}

Based on the best observations of extragalactic H~{\sc{ii}} regions of low
metallicity available in the literature we redetermine the primordial helium
abundance, $Y_p$(nHc), taking into account all the observed He~{\sc{i}} line
intensities with the exception of those strongly affected by underlying
absorption. We derive He/H, $N_e$(He~{\sc{ii}}), $T$(He~{\sc{ii}}), and
$\tau$(3889) self-consistently.

{From} the same data for NGC~2363, SBS~0335--052, Haro~29, and I~Zw~18 we
obtain $Y$(nHc) values about 0.007 smaller than those derived by Izotov, Thuan
and collaborators, the differences are small but significant and are mainly due
to the lower $T$(He~{\sc{ii}}) values adopted by us. In the self-consistent
solutions the lower $T$(He~{\sc{ii}}) values imply higher densities; the higher
the density, the higher the collisional contribution to the He~{\sc{i}} line
intensities and consequently the lower the helium abundances (see Tables
\ref{tchi} and \ref{t346}-\ref{t0335}). The $Y_p$(nHc) value derived by us from
NGC~346, NGC~2363, SBS~0335--052, Haro~29, and I~Zw~18 is in good agreement
with that derived from NGC~346 in Paper~I.

The $Y$(nHc) values are lower limits to the real $Y$ values because they do not
consider the excitation of the Balmer lines due to collisions.  A preliminary
estimate of $Y$(+Hc) is presented in Table~\ref{tcomp}. In a future paper we
will give a full discussion of this problem \citep{lur01} based on detailed
models for each object.

The $Y_p$(nHc) of $0.2356 \pm 0.0020 (1\sigma)$ combined with standard Big Bang
nucleosynthesis computations \citep{tho94,fio98} implies that, at the $2\sigma$
confidence level, $\Omega_b$ is in the 0.011 to 0.020 range for $h$ = 0.7
(see Table~\ref{tbar}).  Also in Table~\ref{tbar} we present seven other
determinations of $\Omega_b$ not based on $Y_p$, our result is in good
agreement with five of them, those based on: the primordial Lithium
determination, Li$_p$, the Cosmic Background Imager, the baryon budget in the
galactic vicinity ($z$ = 0), the baryon budget at $z \sim$ 3.00 and that
derived under the assumption of a flat universe, the SNIa magnitude redshift
data to derive $\Omega_M$, and a baryon fraction estimate based on X-ray
observations of rich clusters of galaxies (this determination is labeled SNIa
in Table~\ref{tbar}); on the other hand our result is in disagreement with two
of them, those based on: the primordial deuterium determination, D$_p$, and
{\sc Boomerang}.

Our $Y_p$(+Hc) value of $0.2384 \pm 0.0025$ is also in good agreement
with the five determinations of $\Omega_b$ mentioned above and is closer than
the $Y_p$(nHc) value to the other two determinations but still in disagreement
with them. 

The errors in all determinations need to be reduced to constrain
even further the domain available to non standard BBN models.

\bigskip

It is a pleasure to acknowledge several fruitful discussions with: G. Ferland,
B. E. J. Pagel, E. Skillman, G. Stasi\'nska, G. Steigman, S. Torres-Peimbert,
and S. Viegas.

\appendix
\begin{appendix}

\section{Second order approximation to the two temperature scheme}

In this appendix we prove the validity of equation~\ref{eTO2+3_T0m}. To
do this, we need to give a physical meaning to 
$T({\rm O~{\scriptstyle{II+III}}})$ and consequently to $\Gamma$, presented in
equation~\ref{eTO2+3def1}. Therefore we need to derive the functional form of
$\Gamma$ that will make equation~\ref{eTO2+3_T0m} accurate to second order.

Equation~\ref{eTO3-T0} is designed (and by analogy \ref{eTO3-T03} and
\ref{eTO2-T02}) to take into account the increase in emissivity of the forbidden
O~{\sc{iii}} lines with increasing temperature, therefore giving higher $T({\rm
O~{\scriptstyle{III}}})-T_0$ differences to regions with larger temperature
fluctuations. We want to give the same physical meaning to
equation~\ref{eTO2+3_T0m}, thus we want the $T({\rm
O~{\scriptstyle{II+III}}})-T_0$ difference to increase with $t^2$.

This can be physically explained by examining equation~\ref{et2} where the
three terms are positive and thus increase $t^2$. The first term is related to
the temperature difference between the O~{\sc{ii}} and O~{\sc{iii}} zones and
the last two terms are related to the temperature fluctuations in each zone.
The fluctuations associated with the last two terms correspond to the $T({\rm
O~{\scriptstyle{II}}})-T_{02}$ and $T({\rm O~{\scriptstyle{III}}})-T_{03}$
Gw3o3h)W
differences, therefore by adopting a $T({\rm O~{\scriptstyle{II+III}}})$ of the
form $(\alpha) T({\rm O~{\scriptstyle{II}}})+(1-\alpha)T({\rm
O~{\scriptstyle{III}}})$ the fluctuations will produce a $T({\rm
O~{\scriptstyle{II+III}}}) - T_0$ difference. The first term should also
contribute to the $T({\rm O~{\scriptstyle{II+III}}})-T_0$ difference, but since
this term does not represent a change in $t_2^2$ or $t_3^2$, $T({\rm
O~{\scriptstyle{II}}})$ and $T({\rm O~{\scriptstyle{III}}})$ will not change
either, consequently another term is needed in the definition of $T({\rm
O~{\scriptstyle{II+III}}})$ This effect is introduced by multiplying the
direct average by a factor $\Gamma$, which represents the second order
correction due to the difference in temperature between the O~{\sc{ii}} and
O~{\sc{iii}} zones, thus $T({\rm O~{\scriptstyle{II+III}}})$ will be of the
form:
\begin{equation}
T({\rm O~{\scriptstyle{II+III}}})= \left[\left(1-\alpha\right)T({\rm
O~{\scriptstyle{III}}}) +\alpha T({\rm O~{\scriptstyle{II}}})\right]\Gamma
\label{eTO2+3def2}
,\end{equation}
where
\begin{equation}
\Gamma=1+\delta^2\gamma.
\label{eGamma}
\end{equation}
Here $\delta$ is the relative difference between $T({\rm
O~{\scriptstyle{II}}})$ and $T({\rm O~{\scriptstyle{III}}})$ and $\gamma$
represents a (free) parameter which will permit to recover the physical meaning
of equations \ref{eTO3-T03} and \ref{eTO2-T02}. Thus $\delta$ is given by
\begin{equation}
\delta=\frac{T({\rm O~{\scriptstyle{III}}})}{T({\rm O~{\scriptstyle{II}}})}-1
\label{edelta_1}
.\end{equation}
To have a second order correction in $\beta$ it is enough to
determine $\delta$ to first order; to first order $T({\rm
O~{\scriptstyle{III}}}) = T_{03}$ and $T({\rm
O~{\scriptstyle{II}}}) = T_{02}$, therefore
\begin{equation}
\delta=\frac{\beta}{\alpha}
\label{edelta_2}
.\end{equation}

Substituting equations \ref{eTO3-T03} and \ref{eTO2-T02} into
equation~\ref{eTO2+3def2} we obtain
\begin{eqnarray}
\frac{T({\rm O~{\scriptstyle{II+III}}})}{T_0} & = &
\Gamma\left[\left(1-\alpha\right)\left(1+\beta\right)
		\left(1-3\frac{t_3^2}{2}\right)+
		\left(1-\alpha\right)\frac{90800}{T_0}\frac{t_3^2}{2}
		\right. \nonumber \\
&&	+ \left.\alpha\left(1+\beta-\frac{\beta}{\alpha}\right)
		\left(1-3\frac{t_2^2}{2}\right)+
		\alpha\frac{97300}{T_0}\frac{t_2^2}{2}\right] \nonumber \\
& = & \Gamma\left[1
		- 3 \left(\left(1-\alpha\right)\frac{t_3^2}{2}+
		\alpha\frac{t_2^2}{2}\right) \right. \nonumber \\
&&	+\left.\frac{1}{T_0}\left(\left(1-\alpha\right)90800\frac{t_3^2}{2}
		+\alpha97300\frac{t_2^2}{2}\right)\right]
\label{eTO2+3_1}
;\end{eqnarray} substituting $\Gamma$ and keeping this equation accurate to
second order in the temperature variations ($\beta$, $t_2$, $t_3$, and $t$) we
obtain
\begin{eqnarray}
\frac{T({\rm O~{\scriptstyle{II+III}}})}{T_0} & = &
	1+\delta^2\gamma \nonumber \\
&&		- 3 \left(\left(1-\alpha\right)t_3^2+
		\alpha t_2^2\right) \frac{1}{2} \nonumber \\
&&	+\frac{1}{T_0}\left(\left(1-\alpha\right)90800t_3^2
		+\alpha 97300 t_2^2\right)\frac{1}{2} 
\label{eTO2+3_2}
.\end{eqnarray}

To simplify this equation and recover the physical meaning of equation
\ref{eTO3-T0} it is necessary for $\gamma$ to be
\begin{equation}
\gamma=\frac{1}{2}\left(1-\alpha\right)\alpha
	\left(\frac{\theta}{T_0}-3\right)
\label{egamma}
,\end{equation}
here it is enough to derive $\gamma$ to zeroth order, so dividing $\theta$ by
any one of the available temperatures ($T_0$, $T_{02}$, $T_{03}$, $T({\rm
O~{\scriptstyle{II}}})$ or $T({\rm O~{\scriptstyle{III}}})$) would be
approximately right, thus we obtain
\begin{eqnarray}
\frac{T({\rm O~{\scriptstyle{II+III}}})}{T_0} & = &
	1 - 3 \left(\left(1-\alpha\right)t_3^2+\alpha t_2^2+
		\beta^2\frac{1-\alpha}{\alpha}\right)\frac{1}{2} \nonumber \\
&&	+\frac{90800}{T_0}\left(1-\alpha\right)\left(t_3^2+
		\beta^2\frac{1-\alpha}{\alpha}\right)\frac{1}{2} \nonumber \\
&&	+\frac{97300}{T_0}\alpha\left(t_2^2+
		\beta^2\frac{1-\alpha}{\alpha}\right)\frac{1}{2}
\label{eTO2+3_3}
.\end{eqnarray}

Under the assumption that
\begin{equation}
t_2^2 \approx t_3^2
\label{et22t32}
,\end{equation}
equation~\ref{et2} becomes 
\begin{equation}
t^2 = t_2^2 + \beta^2\frac{1-\alpha}{\alpha}
\label{et2fin}
,\end{equation}
and finally equation~\ref{eTO2+3_3} becomes
\begin{equation}
T({\rm O~{\scriptstyle{II+III}}})=T_0
\left( 1 + \left[\frac{\theta}{T_0}-3 \right] \frac{t^2}{2} \right)
\label{eTO2+3_T0d}
,\end{equation}
or
\begin{equation}
T_0 = T({\rm O~{\scriptstyle{II+III}}})
		\left(1-
		\left[\frac{\theta}{T({\rm O~{\scriptstyle{II+III}}})}
		-3\right]
		\frac{t^2}{2}\right)
\label{eT0-TO2+3}
.\end{equation}
Equation~\ref{eT0-TO2+3} can now be combined with those equations that relate
$T_0$ and $t^2$, to some observable temperature like $T$(Bac) or $T({\rm
He~{\scriptstyle{II}}})$.

To adopt a value of $\Gamma=1$ in equation~\ref{eTO2+3def1} is equivalent to
assume that the first term in equation~\ref{et2} is equal to zero. We have
already mentioned the parameter space where it is a good approximation to
assume $\Gamma=1$. To obtain very accurate He/H values the exact value of
$\Gamma$, that includes the second order effects implied by $t^2$, should be
used. 

\end{appendix}

\clearpage

\begin{deluxetable}{lr@{$\pm$}lr@{$\pm$}lr@{$\pm$}lr@{$\pm$}lr@{$\pm$}l}
\tablecaption{Adopted He~{\sc{i}} line intensities relative to H$\beta$
\label{tlines}}
\tablewidth{0pt}
\tablehead{
\colhead{He~{\sc{i}} line}&
\multicolumn{2}{c}{NGC~346}&
\multicolumn{2}{c}{I~Zw~18}&
\multicolumn{2}{c}{NGC~2363}&
\multicolumn{2}{c}{Haro~29}&
\multicolumn{2}{c}{SBS~$0335-052$}}
\startdata
3820&
	\multicolumn{2}{c}{...}&
		\multicolumn{2}{c}{...}&
			0.012&0.001\tablenotemark{c}&
				0.010&0.001\tablenotemark{c}&
					\multicolumn{2}{c}{...}\\
3889\tablenotemark{a}&
	0.0940&0.0017\tablenotemark{b}
		& 0.0898&0.0071\tablenotemark{c}&
			0.087&0.001\tablenotemark{c}&
				0.095&0.001\tablenotemark{c}&
					0.0924&0.0028\tablenotemark{c,f}\\
4026&
	0.0185&0.0007\tablenotemark{b}&
		0.0202&0.0036\tablenotemark{c}&
			0.018&0.001\tablenotemark{c}&
				0.020&0.001\tablenotemark{c}&
					0.0173&0.0006\tablenotemark{c,f}\\
4387&
	0.0047&0.0002\tablenotemark{b}&
		\multicolumn{2}{c}{...}&
			0.005&0.001\tablenotemark{c}&
				0.005&0.001\tablenotemark{c}&
					\multicolumn{2}{c}{...}\\
4471&
	0.0384&0.0005\tablenotemark{b}&
		0.0387&0.0025\tablenotemark{c}&
			0.040&0.001\tablenotemark{c}&
				0.039&0.001\tablenotemark{c}&
					0.0382&0.0007\tablenotemark{c,f}\\
4922&
	0.0100&0.0002\tablenotemark{b}&
		\multicolumn{2}{c}{...}&
			0.013&0.001\tablenotemark{c}&
				\multicolumn{2}{c}{...\tablenotemark{e}}&
					0.0093&0.0004\tablenotemark{c,f}\\
5876&
	0.1064&0.0012\tablenotemark{b}&
		0.0936&0.0028\tablenotemark{d}&
			0.106&0.001\tablenotemark{d}&
				0.102&0.001\tablenotemark{d}&
					0.1027&0.0017\tablenotemark{f}\\
6678&
	0.0296&0.0002\tablenotemark{b}&
		0.0263&0.0018\tablenotemark{d}&
			0.029&0.001\tablenotemark{d}&
				0.029&0.001\tablenotemark{d}&
					0.0263&0.0006\tablenotemark{f}\\
7065&
	0.0211&0.0002\tablenotemark{b}&
		0.0239&0.0016\tablenotemark{d}&
			0.029&0.001\tablenotemark{d}&
				0.024&0.001\tablenotemark{d}&
					0.0358&0.0007\tablenotemark{f}\\
7281&
	0.0063&0.0003\tablenotemark{b}&
		\multicolumn{2}{c}{...}&		
			0.006&0.001\tablenotemark{d}&
				0.005&0.001\tablenotemark{d}&
					\multicolumn{2}{c}{...}\\
\enddata

\tablenotetext{a}{Where the contribution due to the H8 line has been
subtracted.}
\tablenotetext{b}{From \citet{pei00z}.}
\tablenotetext{c}{From \citet{izo97,izo99}, but modified due to underlying
absorption, see text.}
\tablenotetext{d}{From \citet{izo97,izo99}.}
\tablenotetext{e}{Presented by \citet{izo97}, but not used, see text.}
\tablenotetext{f}{Modified by assuming a different H$\beta$ underlying
absorption from that used by \citet{izo99}, see text.}
\end{deluxetable}

\clearpage

\begin{deluxetable}{lccccccccc}
\tablecaption{$N({\rm He}^+)/N({\rm H}^+)$\tablenotemark{a} \
and $\chi^2$ for Region A
\label{tchi}}
\tablewidth{0pt}
\tablehead{
&&&&\multicolumn{6}{c}{$N_e$(He~{\sc{ii}})} \\
&&&&\multicolumn{6}{c}{(cm$^{-3}$)} \\
&& \colhead{$T$(He~{\sc{ii}}) } & \ & \cline{1-6} \\
\colhead{$t^2$} & \ & \colhead{(K)} & \ & 
\colhead{69}  & \colhead{90}  & \colhead{110} &
\colhead{144} & \colhead{182} & \colhead{220}}
\startdata
0.001\tablenotemark{b}
     &&12,800&& 8052                    & 8014 & 7980 & 7924 & 7865 & 7810 \\
     &&      &&(12.4)\tablenotemark{c}  &(14.0)&(18.4)&(31.7)&(54.2)&(83.3) \\
\\
0.014&&12,240&& 8036 & 8001 & 7969                    & 7917 & 7861 & 7809 \\
     &&      &&(13.0)&(8.81)&(7.53)\tablenotemark{c}  &(10.8)&(21.7)&(39.0) \\
\\
0.022&&11,860&& 8026 & 7992 & 7961 & 7911                    & 7858 & 7808 \\
     &&      &&(22.3)&(14.5)&(9.58)&(6.53)\tablenotemark{c,d}&(10.0)&(19.8) \\
\\
0.030&&11,480&& 8015 & 7983 & 7953 & 7906 & 7855                    & 7807 \\
     &&      &&(39.0)&(27.9)&(19.7)&(10.8)&(7.53)\tablenotemark{c}  &(10.3) \\

\enddata
\tablenotetext{a}{Given in units of $10^{-5}$, for the case of
no hydrogen collisions, $\chi^2$ values in parenthesis.}
\tablenotetext{b}{Minimum $t^2$ value, assuming that 
$t_{2}{^2} = t_{3}{^2} = 0.000$ and considering the difference between $T_{02}$
and $T_{03}$, see text.}
\tablenotetext{c}{The minimum $\chi^2$ value at a given $t^2$.}
\tablenotetext{d}{The smallest $\chi^2$ value for all $t^{2}$'s and
densities, thus defining $T$(He~{\sc{ii}}) and $N_e$(He~{\sc{ii}}).}
\end{deluxetable}

\clearpage

\begin{deluxetable}{lccccc}
\tablecaption{Parameters of the preferred model for each H~{\sc{ii}} region.
\label{tparameters}}
\tablewidth{0pt}
\tablehead{
\colhead{Parameter}
&\colhead{NGC~346}
&\colhead{I~Zw~18}
&\colhead{NGC~2363}
&\colhead{Haro~29}
&\colhead{SBS~$0335-052$}}
\startdata
$T$(O~{\sc{iii}})\tablenotemark{a}&
	$13070\pm50$&
		$19060\pm610$&
			$15750\pm100$&
				$16050\pm100$&
					$20500\pm200$\\
$T$(He~{\sc{ii}})\tablenotemark{b}&
	$11860\pm370$&
		$17120\pm600$&
			$14710\pm440$&
				$14880\pm300$&
					$19290\pm310$\\
$t^2$\tablenotemark{b}&
	$0.022\pm0.008$&
		$0.024\pm0.006$&
			$0.021\pm0.010$&
				$0.023\pm0.007$&
					$0.021\pm0.007$\\
$N$(He~{\sc{ii}})\tablenotemark{b}&
	$144^{+44}_{-38}$&
		$87^{+94}_{-78}$&
			$250^{+88}_{-77}$&
				$235\pm85$\tablenotemark{c}&
					$297^{+65}_{-56}$\\
$\tau(3889)$\tablenotemark{b}&
	$0.000$&
		$0.010\pm0.005$&
			$0.98\pm0.25$&
				$0.41\pm0.30$&
					$1.60\pm0.35$\\
$ICF$(He)\tablenotemark{c}&
	1.000&
		1.000&
			0.993&
				0.995&
					0.9985\\
\enddata
\tablenotetext{a}{Observed value.}
\tablenotetext{b}{Output from the MLM fit, see text.}
\tablenotetext{c}{From photoionization models, see text.}
\end{deluxetable}

\clearpage

\begin{deluxetable}{lccccccccc}
\tablecaption{$Y$(nHc) values for NGC~346
\label{t346}}
\tablewidth{0pt}
\tablehead{
&&&&\multicolumn{6}{c}{$N_e$(He~{\sc{ii}})} \\
&&&&\multicolumn{6}{c}{(cm$^{-3}$)} \\
&& \colhead{$T$(He~{\sc{ii}})} & \ & \cline{1-6} \\
\colhead{$t^2$} & \ & \colhead{(K)} & \ &
\colhead{69}  & \colhead{90}  & \colhead{110} &
\colhead{144} & \colhead{182} & \colhead{220}}
\startdata
0.001\tablenotemark{a}
     &&12,800&&0.2436\tablenotemark{b}  &0.2428&0.2420&0.2407&0.2393&0.2381 \\
0.014&&12,240&&0.2433&0.2425&0.2417\tablenotemark{b}  &0.2405&0.2392&0.2380 \\
0.022&&11,860&&0.2430&0.2423&0.2415&0.2404\tablenotemark{b,c}&0.2392&0.2380 \\
0.030&&11,480&&0.2428&0.2420&0.2414&0.2403&0.2391\tablenotemark{b}  &0.2380 \\

\enddata
\tablenotetext{a}{Minimum $t^2$ value, assuming that 
$t_{2}{^2} = t_{3}{^2} = 0.000$ and considering the difference between $T_{02}$
and $T_{03}$, see text.}
\tablenotetext{b}{Minimum $\chi^2$ value at a given $t^2$,
see Table~\ref{tchi}.}
\tablenotetext{c}{The smallest $\chi^2$ value for this object,
see Table~\ref{tchi}.} 
\end{deluxetable}

\clearpage

\begin{deluxetable}{lccccccccc}
\tablecaption{$Y$(nHc) values for I~Z{\protect\MakeLowercase{w}}~18
\label{tZw}}
\tablewidth{0pt}
\tablehead{
&&&&\multicolumn{6}{c}{$N_e$(He~{\sc{ii}})} \\
&&&&\multicolumn{6}{c}{(cm$^{-3}$)} \\
&& \colhead{$T$(He~{\sc{ii}})} & \ & \cline{1-6} \\
\colhead{$t^2$} & \ & \colhead{(K)} & \ &
\colhead{25}  & \colhead{59}  & \colhead{77} &
\colhead{87}  & \colhead{97} & \colhead{150}}
\startdata
0.006\tablenotemark{a}
     &&17,920&&0.2404&0.2375\tablenotemark{b}  &0.2362&0.2354&0.2346&0.2307 \\
0.018&&17,390&&0.2400&0.2373&0.2359\tablenotemark{b}  &0.2351&0.2344&0.2307 \\
0.024&&17,120&&0.2398&0.2371&0.2358&0.2350\tablenotemark{b,c}&0.2343&0.2307 \\
0.030&&16,860&&0.2396&0.2370&0.2357&0.2349&0.2342\tablenotemark{b}  &0.2307 \\
    
\enddata
\tablenotetext{a}{Minimum $t^2$ value, assuming that 
$t_{2}{^2} = t_{3}{^2} = 0.000$ and considering the difference between $T_{02}$
and $T_{03}$, see text.}
\tablenotetext{b}{Minimum $\chi^2$ value at a given $t^2$.}
\tablenotetext{c}{Best fit (minimum $\chi^2$ value) for this object.} 

\end{deluxetable}

\clearpage

\begin{deluxetable}{lccccccccc}
\tablecaption{$Y$(nHc) values for NGC~2363
\label{t2363}}
\tablewidth{0pt}
\tablehead{
&&&&\multicolumn{6}{c}{$N_e$(He~{\sc{ii}})} \\
&&&&\multicolumn{6}{c}{(cm$^{-3}$)} \\
&& \colhead{$T$(He~{\sc{ii}})} & \ & \cline{1-6} \\
\colhead{$t^2$} & \ & \colhead{(K)} & \ &
\colhead{100} & \colhead{183} & \colhead{214} &
\colhead{250} & \colhead{289} & \colhead{338}}
\startdata
0.001\tablenotemark{a}
     &&15,620&&0.2500&0.2459\tablenotemark{b}  &0.2445&0.2430&0.2414&0.2395 \\
0.011&&15,160&&0.2493&0.2455&0.2442\tablenotemark{b}  &0.2427&0.2412&0.2394 \\
0.021&&14,710&&0.2487&0.2451&0.2438&0.2424\tablenotemark{b,c}&0.2410&0.2393 \\
0.031&&14,260&&0.2480&0.2446&0.2434&0.2421&0.2407\tablenotemark{b}  &0.2391 \\
     
\enddata
\tablenotetext{a}{Minimum $t^2$ value, assuming that 
$t_{2}{^2} = t_{3}{^2} = 0.000$ and considering the difference between $T_{02}$
and $T_{03}$, see text.}
\tablenotetext{b}{Minimum $\chi^2$ value at a given $t^2$.}
\tablenotetext{c}{Best fit (minimum $\chi^2$ value) for this object.} 
\end{deluxetable}

\clearpage

\begin{deluxetable}{lccccccccc}
\tablecaption{$Y$(nHc) values for Haro~29
\label{th29}}
\tablewidth{0pt}
\tablehead{
&&&&\multicolumn{6}{c}{$N_e$(He~{\sc{ii}})} \\
&&&&\multicolumn{6}{c}{(cm$^{-3}$)} \\
&& \colhead{$T$(He~{\sc{ii}})} & \ & \cline{1-6} \\
\colhead{$t^2$} & \ & \colhead{(K)} & \ &
\colhead{50}  & \colhead{100} & \colhead{150} &
\colhead{235} & \colhead{320} & \colhead{500}}
\startdata
0.001\tablenotemark{a}
     &&15,860&&0.2477&0.2451&0.2427&0.2390                 &0.2356&0.2296 \\
0.016&&15,190&&0.2471&0.2446&0.2424&0.2389                 &0.2356&0.2200 \\
0.023&&14,880&&0.2468&0.2444&0.2422&0.2388\tablenotemark{b}&0.2356&0.2302 \\
0.030&&14,570&&0.2465&0.2441&0.2421&0.2388                 &0.2356&0.2304 \\
    
\enddata
\tablenotetext{a}{Minimum $t^2$ value, assuming that 
$t_{2}{^2} = t_{3}{^2} = 0.000$ and considering the difference between $T_{02}$
and $T_{03}$, see text.}
\tablenotetext{b}{Preferred value, see text.} 
\end{deluxetable}

\clearpage

\begin{deluxetable}{lccccccccc}
\tablecaption{$Y$(nHc) values for SBS~0335--052
\label{t0335}}
\tablewidth{0pt}
\tablehead{
&&&&\multicolumn{6}{c}{$N_e$(He~{\sc{ii}})} \\
&&&&\multicolumn{6}{c}{(cm$^{-3}$)} \\
&& \colhead{$T$(He~{\sc{ii}})} & \ & \cline{1-6} \\
\colhead{$t^2$} & \ & \colhead{(K)} & \ &
\colhead{241} & \colhead{257} & \colhead{279} &
\colhead{297} & \colhead{316} & \colhead{362}}
\startdata
0.004\tablenotemark{a}
     &&20,020&&0.2439&0.2425\tablenotemark{b}  &0.2409&0.2396&0.2382&0.2351 \\
0.014&&19,590&&0.2442&0.2430&0.2413\tablenotemark{b}  &0.2400&0.2387&0.2356 \\
0.021&&19,290&&0.2445&0.2433&0.2416&0.2403\tablenotemark{b,c}&0.2390&0.2360 \\
0.028&&18,990&&0.2447&0.2435&0.2419&0.2406&0.2393\tablenotemark{b}  &0.2363 \\
     
\enddata

\tablenotetext{a}{Minimum $t^2$ value, assuming that 
$t_{2}{^2} = t_{3}{^2} = 0.000$ and considering the difference between $T_{02}$
and $T_{03}$, see text.}
\tablenotetext{b}{Minimum $\chi^2$ value at a given $t^2$.}
\tablenotetext{c}{Best fit (minimum $\chi^2$ value) for this object.} 
\end{deluxetable}

\clearpage
\begin{deluxetable}{lcccc}
\tablecaption{$Y$ Comparison
\label{tcomp}}
\tablewidth{0pt}
\tablehead{ && \colhead{$Y$(nHc)\tablenotemark{b}} &
\colhead{$Y$(nHc)\tablenotemark{c}} & \colhead{$Y$(+Hc)\tablenotemark{c}}\\
& \colhead{$O$\tablenotemark{a}} & \colhead{$t^2=0.000$} & \colhead{$t^2 \neq 0.000$}
& \colhead{$t^2 \neq 0.000$}} 
\startdata
NGC~346       & $1714 \pm 255$ &  .....               &
$0.2405 \pm 0.0017$ & $0.2421 \pm 0.0018$ \\
NGC~2363      & $1155 \pm ~97$ &  $0.2456 \pm 0.0008$ &
$0.2424 \pm 0.0035$ & $0.2456 \pm 0.0039$ \\
Haro~29       & $~948 \pm 122$ &  $0.2509 \pm 0.0012$ &
$0.2388 \pm 0.0040$ & $0.2420 \pm 0.0044$ \\
SBS~0335--052 & $~392 \pm ~36$ &  $0.2463 \pm 0.0015$ &
$0.2403 \pm 0.0048$ & $0.2450 \pm 0.0060$ \\
I~Zw~18       & $~243 \pm ~24$ &  $0.2429 \pm 0.0070$ &
$0.2350 \pm 0.0072$ & $0.2412 \pm 0.0085$ \\[2.0ex]

$Y_p$(sample) & & $0.2445 \pm 0.0009$\tablenotemark{d}&
$0.2356 \pm 0.0020$\tablenotemark{d} & $0.2386 \pm 0.0025$\tablenotemark{d} \\
\enddata
\tablenotetext{a}{Oxygen abundance by mass for $t^2 \sim 0.02$, in units of $10^{-6}$.}
\tablenotetext{b}{\citet{izo98,izo99}.} 
\tablenotetext{c}{This paper.} 
\tablenotetext{d}{Derived from the sample under the assumption that 
$\Delta Y/\Delta O = 3.5 \pm 0.9$ (see Paper~I).}
\end{deluxetable}

\clearpage

\begin{deluxetable}{lcc@{\hspace{36pt}}l}
\tablecaption{Baryon densities for $h=0.70$
\label{tbar}}
\tablewidth{0pt}
\tablehead{\colhead{Method} & \colhead{$\Omega_b$} && \colhead{Source}}
\startdata
D$_p$                                         &  0.035 - 0.049 & ($2\sigma$) & \citet{ome01}  \\
Li$_p$                                        &  0.013 - 0.026 & ($2\sigma$) & \citet{suz00}  \\
{\sc Boomerang}                               &  0.031 - 0.059 & ($2\sigma$) & \citet{net01} \\
CBI                                           &  0.006 - 0.060 & (min-max)   & \citet{pad01}  \\
($z=0$)                                       &  0.007 - 0.041 & (min-max)   & \citet{fuk98}  \\
($z\sim3.0$)                                  &  0.010 - 0.060 & (min-max)   & \citet{fuk98}  \\
SNIa                                          &  0.018 - 0.066 & ($2\sigma$) & \citet{ste01}  \\[1.5ex]
$Y_p$(nHc),$t^2 = 0.000$                      &  0.025 - 0.037 & ($2\sigma$) & this paper     \\
$Y_p$(nHc),$t^2 \neq 0.000$ \tablenotemark{a} &  0.011 - 0.020 & ($2\sigma$) & this paper     \\
$Y_p$(+Hc),$t^2 \neq 0.000$ \tablenotemark{a} &  0.012 - 0.027 & ($2\sigma$) & this paper     \\
\enddata
\tablenotetext{a} {Recommended values, see text.}
\end{deluxetable}

\clearpage

\begin{figure}
\begin{center}
\includegraphics[scale=.77]{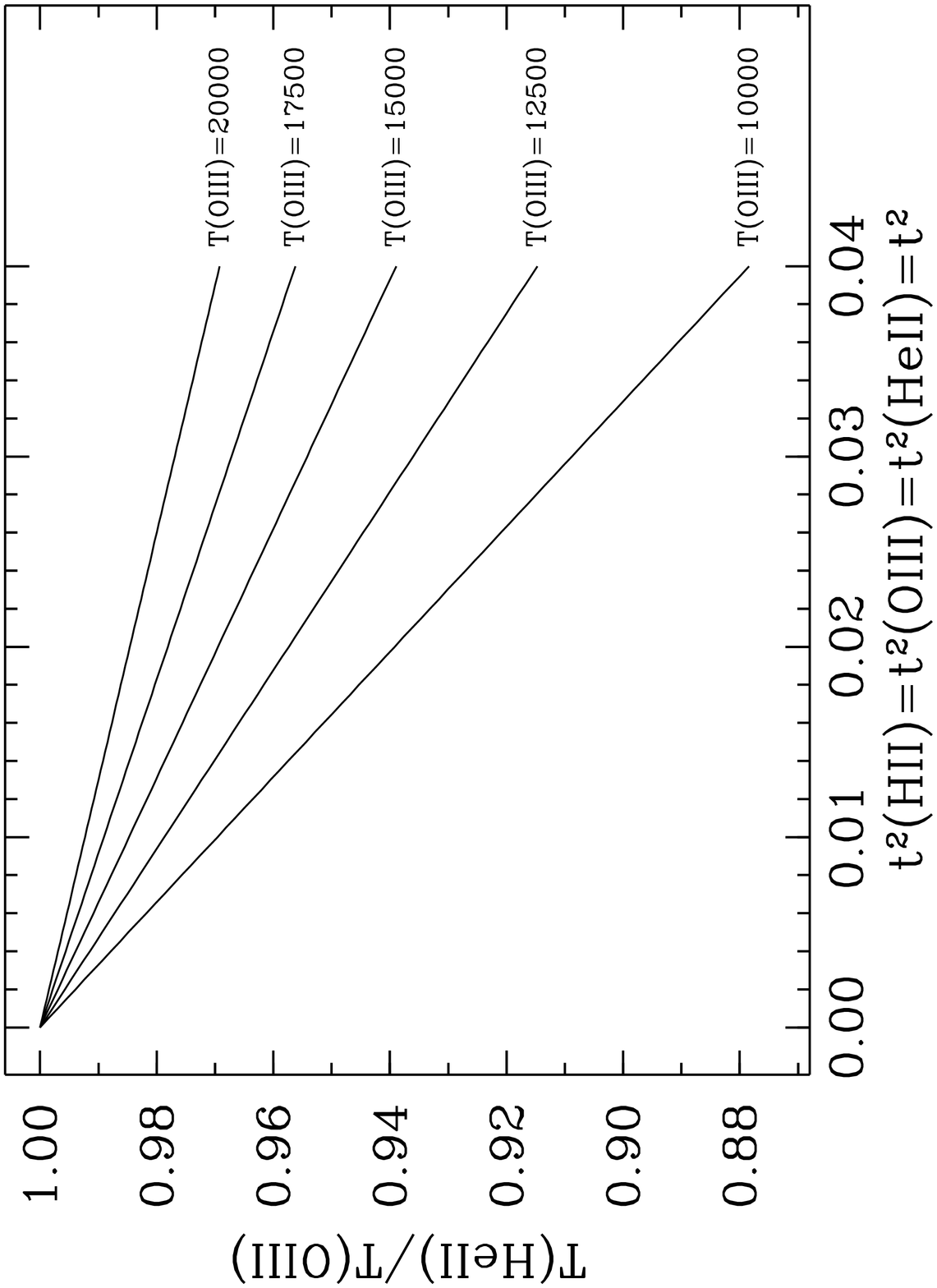}
\end{center}
\end{figure}

\figcaption[f1.eps]{
\label{fTO3-THe}
$T_e$(He~{\sc{ii}})/$T_e$(O~{\sc{iii}}) as a function of $T_e$(O~{\sc{iii}})
and temperature fluctuations for the case in which all the O is O$^{++}$. When
O$^+$ is present higher $t^2$ values are expected, particularly for those
objects with the highest $T_e$(O~{\sc{iii}}) values (see Figure 2). Typical
$t^2$ values in H~{\sc{ii}} regions are in the 0.01 to 0.04 range.
}

\begin{figure}
\begin{center}
\includegraphics[scale=.77]{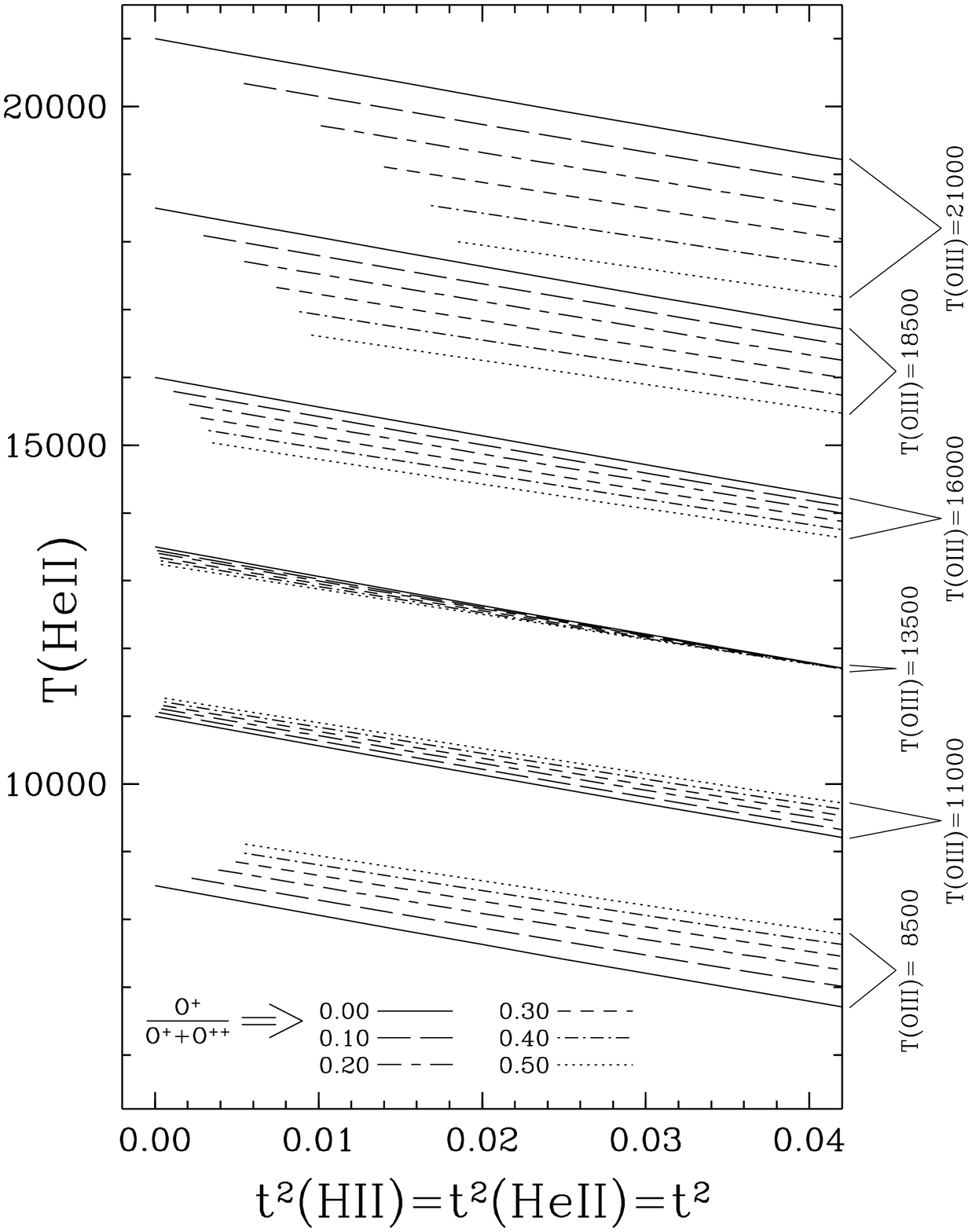}
\end{center}
\end{figure}

\figcaption[f2.eps]{
\label{fTO2+3}
$T_e$(O~{\sc{iii}}) versus $T_e$(He~{\sc{ii}}) showing the effect of different
O$^+$ fractions and different total $t^2$ on $T_e$(He~{\sc{ii}}). Notice that
when O$^+$ is present there are forbidden regions that correspond to the low
end of $t^2$ (those regions where the lines stop) because it is unphysical to
assume a uniform $T_e$(He~{\sc{ii}}) when one is already using 2 different
temperatures for a given line of sight: $T_e$(O~{\sc{iii}}) $and$
$T_e$(O~{\sc{ii}}).  The $T_e$(O~{\sc{ii}}) for the O$^+$ fraction is assumed
to be the one derived from the photoionization models by \citet {sta90}.
}

\begin{figure}
\begin{center}
\includegraphics[scale=.77]{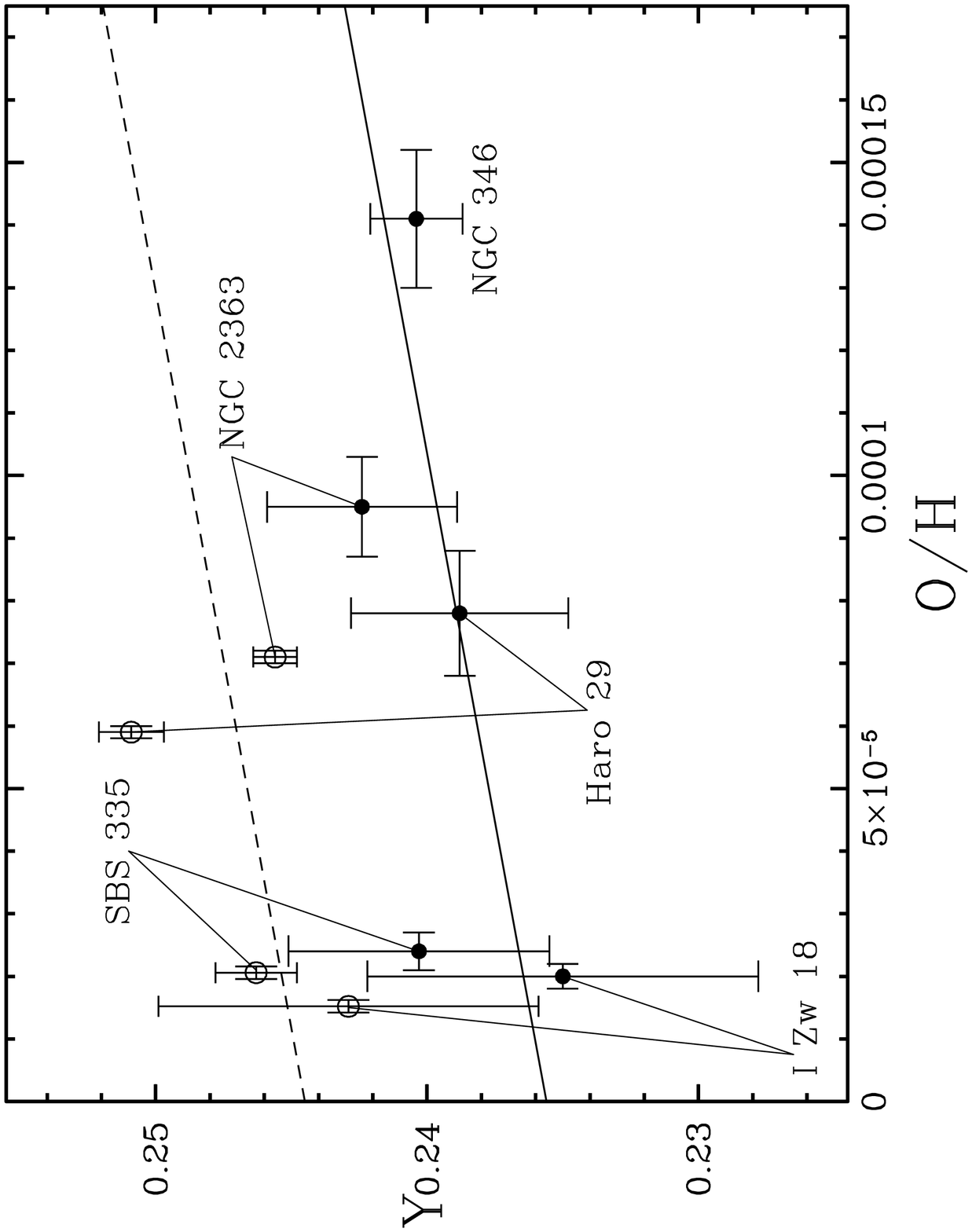}
\end{center}
\end{figure}

\figcaption[f3.eps]{
\label{fY-OH}
$Y$(nHc) versus O/H diagram. Results for $t^2=0.000$ \citep{izo98,izo99} are
shown as open circles, the error bars are those quoted by the authors, they
include the errors in the measurements of the lines, but assume no uncertainties
on the determination of $T_e$(He~{\sc{ii}}), $N_e$(He~{\sc{ii}}), or
$\tau(3889)$; results for $t^2 \neq 0.000$ (this paper) are shown as solid
dots, the errors represent the uncertainties on: the intensities of the lines,
$T_e$(He~{\sc{ii}}), $N_e$(He~{\sc{ii}}), and $\tau(3889)$; other possible
systematic effects (such as the collisional contribution to the Balmer lines,
or errors on the atomic physics) are not included, see text. The lines
represent the best fits to the data assuming a slope of $\Delta Y / \Delta O =
3.5 \pm 0.9$, see text.  }

\begin{figure}
\begin{center}
\includegraphics[scale=.77]{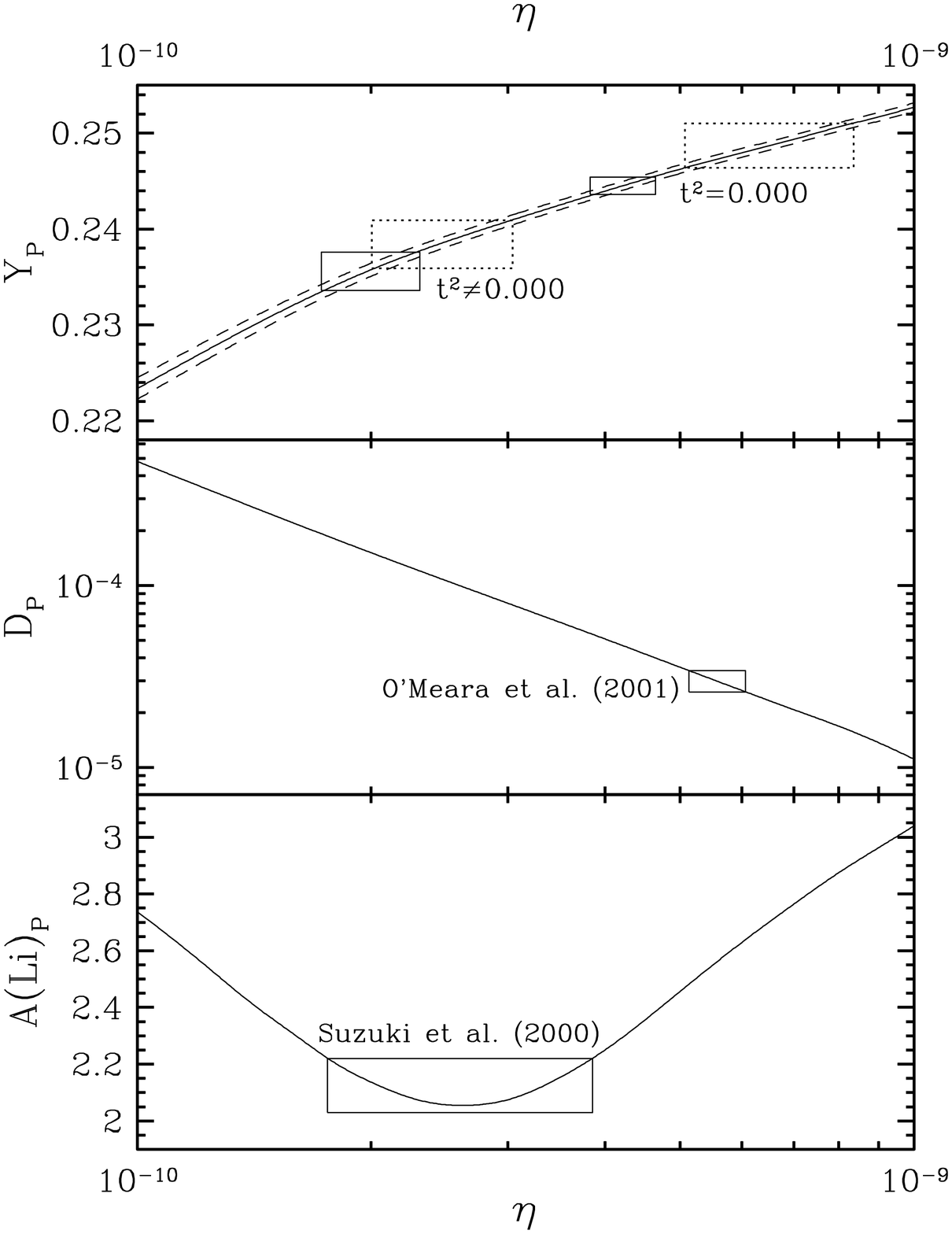}
\end{center}
\end{figure}

\figcaption[f4.eps]{
\label{feta}
Helium, deuterium, and lithium abundances predicted by standard Big Bang
nucleosynthesis computations with three light neutrino species for different
values of $\eta$, the baryon to photon ratio \citep{tho94,fio98}.  Also in this
figure we present observational abundance determinations of these elements. The
helium boxes are $1\sigma$ determinations from this work; the solid line boxes
correspond to $Y_p$(nHc) and the dotted line boxes to $Y_p$(+Hc) (see Tables
\ref{tcomp}~and~\ref{tbar}). The deuterium box is the $1\sigma$ determination
by \citet{ome01} while the lithium box is the $2\sigma$ determination by
\citet*{suz00}.
}
\end{document}